\documentclass[preprint,showpacs]{revtex4-1}
\usepackage[pdftex]{hyperref,graphicx}	
\usepackage[loose]{units}
\usepackage{amsmath}
\usepackage{amssymb}
\usepackage[utf8]{inputenc}
\usepackage[section]{placeins}
\usepackage{natbib}

\usepackage{color} 
\usepackage[normalem]{ulem} 

\newcommand{\m}[1]{\mathrm{#1}}

\begin{document}

\title{Diffusion and bulk flow in phloem loading:\\ A theoretical analysis of the polymer trap mechanism for sugar transport in plants}

\author{Julia D\"olger$^{1,3}$, Hanna Rademaker$^1$, Johannes Liesche$^2$, Alexander Schulz$^2$ and Tomas Bohr$^1$} 
\affiliation{$^1$Department of Physics and Center for Fluid Dynamics, Technical University of Denmark, Kgs. Lyngby, Denmark\\
$^2$ Department of Plant and Environmental Sciences, University of Copenhagen, Copenhagen, Denmark\\
$^3$ Institute for Condensed Matter Physics, Darmstadt University of Technology, Darmstadt, Germany
}

\vspace{5mm} 

\begin{abstract}
Plants create sugar in the mesophyll cells of their leaves by photosynthesis. This sugar, mostly sucrose, has to be {\em loaded} via the bundle sheath into the phloem vascular system (the sieve elements), where it is distributed to growing parts of the plant. We analyze the feasibility of a particular loading mechanism, {\em active symplasmic loading}, also called the {\em polymer trap} mechanism, where sucrose  is  transformed into heavier sugars, such as raffinose and stachyose, in the intermediary-type companion cells bordering the sieve elements in the minor veins of the phloem. Keeping the heavier sugars from diffusing back requires that the {\em plasmodesmata} connecting the bundle sheath with the intermediary cell act as extremely precise filters, which are able to distinguish between molecules that differ by less than 20\% in size.  In our modeling, we take into account the coupled water and sugar movement across the relevant interfaces, without explicitly considering the chemical reactions transforming the sucrose into the heavier sugars. Based on the available data for plasmodesmata geometry, sugar concentrations and flux rates, we conclude that this mechanism can in principle function, but that it requires pores of molecular sizes. Comparing with the somewhat uncertain experimental values for sugar export rates, we expect the pores to be only 5-10\% larger than the hydraulic radius of the sucrose molecules. We find that the water flow through the plasmodesmata, which has not been quantified before, contributes only 10-20\% to the sucrose flux into the intermediary cells, while the main part is transported by diffusion. On the other hand, the subsequent sugar translocation into the sieve elements would very likely be carried predominantly by bulk water flow through the plasmodesmata.
Thus, in contrast to apoplasmic loaders, all the necessary water for phloem translocation would be supplied in this way with no need for additional water uptake across the plasma membranes of the phloem. 
 
\begin{center}
\today
\end{center}
\end{abstract}

\maketitle

\section{Introduction}
Leaves maintain an extremely delicate balance between water and sugar translocation to ensure the outflow and eventual evaporation of water from the {\em xylem} cells simultaneously with the inflow of water and sugar to the {\em phloem} cells nearby. Xylem and phloem are the two long distance pathways in vascular plants, where the former conducts water from the roots to the leaves and the latter distributes the sugar produced in the leaves. The sugar which is loaded into the {\em sieve elements}, the conducting cells of the phloem is generated in the chloroplasts of the {\em mesophyll cells} outside the {\em bundle sheath}, a layer of tightly arranged cells around the vascular bundle, which protects the veins of both xylem and phloem from the air present in the space between the mesophyll cells and the {\em stomata}. The latter are specialised cells, that control the air flow in and out of the leaf by adjusting the size of pores in the epidermis. The water which leaves the xylem is under negative pressure, up to \(-80\) bars have been reported 
\cite{Scholander1965}, 
whereas the water in the phloem a few micrometers away is under positive pressure, typically around \(+10\) bars \cite{Taiz}. On the other hand, the sugar concentration is close to 0 in the xylem and up to 1 molar in the phloem, where the {\em M{\"u}nch mechanism} \cite{munch1930} is believed to be responsible for the flow: the large sugar concentrations in the phloem cells of the mature  ``source" leaves will by osmosis increase the pressure and drive a bulk flow towards the various ``sinks", where sugar is used.  

The water flow from the xylem has two important goals: most of it evaporates, presumably from the walls of the mesophyll cells, maintaining the negative pressures in the xylem necessary to draw water from the roots, but a small part of it passes across the plasma membranes into the mesophyll cells and takes part in the photosynthesis and the subsequent translocation of the sugars through the bundle sheath towards the sieve elements of the phloem. This loading process is not understood in detail, but several important characteristics are known and plants have been divided into rough categories \cite{rennie2009a} depending on their loading mechanisms. Many trees
are so-called ``passive loaders", which means that the sugar concentration is largest in the mesophyll and decreases towards the sieve cells. This implies that sugar could simply diffuse from mesophyll cells to sieve elements without any active mechanism.

In other plants the concentrations are reversed, with the largest concentration occurring in the phloem, which then involves some active mechanism. An interesting class of plants is believed to make use of the so-called ``active symplasmic" loading or ``polymer trap" mechanism \cite{rennie2009a}, which is illustrated in Fig. \ref{polymertrap}. Here high concentrations, and thus efficient sugar translocation in the sap, are achieved actively, by transforming the sucrose generated in the mesophyll and transported into the bundle sheath into heavier sugars, the oligosaccharides raffinose and stachyose, which are too large to diffuse back.

The flow into the phloem can follow two pathways, either through the  {\em symplasm} (the interior of the cells) or through the {\em apoplast} (the space outside the plasma membranes, e.g., cell walls). In symplasmic loaders abundant plasmodesmata, i.e., membrane-surrounded channels through the cell walls, provide continuity of the loading pathway and therefore the sugar does not have to pass the plasma membranes as shown in Fig. \ref{polymertrap}. It has recently been pointed out that the polymer trap mechanism would require plasmodesmata with very specific filtering properties allowing sufficient amounts of sucrose to pass while blocking the heavier sugars \cite{liesche2013a}.

We analyze this question in the present paper including both sugar diffusion and bulk water flow in our model without explicitly considering the chemical
reactions transforming the sucrose into the heavier sugars. We restrict the scope of our model to the part of the leaf where the loading of sugar into the phloem transport system takes place. We therefore only include one bundle sheath cell (BSC), intermediary cell (IC) and sieve element (SE) and their interfaces in our study. We also restrict the model to a steady-state situation in which flows, concentrations and pressures are constant. We derive and solve general equations for this setup and check their plausibility and implications with the help of the most complete set of measured values that we could find (for \textit{Cucumis melo}).
The phloem cells in the leaf need water for sugar translocation and they need to build up sufficient pressure ($p_3$ in Fig. \ref{polymertrap}) to generate efficient bulk movement of the phloem sap. On the other hand, the pressure cannot be too high in cells which are exposed to the xylem. Otherwise they would lose water across the water permeable plasma membrane towards the apoplast. If sugar is loaded only via diffusion without any significant water flow, the sieve element has to draw in the water from the surroundings across its plasma membrane. This requires a sufficiently low {\em water potential} $\Psi=p-R T c$ in the phloem, i.e., a hydrostatic pressure $p$ significantly lower than the osmotic pressure $R T c$. If, on the other hand, enough water flows along with the sugar through the plasmodesmata, i.e., symplasmically, the plant does not have to draw in water across the plasma membrane of the phloem cells (sieve element plus intermediary cells) and the hydrostatic pressure can therefore be greater, leading to more efficient vascular flow.
In the following we shall point out a likely scenario (see Sec. \ref{secB}), in which the polymer trap mechanism can function. We stress that this conclusion is based on very limited experimental information. There is a severe lack of precise knowledge on the anatomy of the plasmodesmata, the precise sugar concentrations (taking sufficient account of the distribution of the sugars inside the compartments of the cells) and as the most severe problem, an almost total lack of pressure measurements. The latter reflects the fact that determination of the pressure in a functioning (living) phloem is at present not feasible. \\
From our analysis, however, some important features of this special and fascinating loading mechanism has become clear. Analysing simple equilibrium configurations with the use of irreversible thermodynamics (Kedem-Kachalsky equations) and the theory of hindered transport, we show that diffusion can in fact, despite claims to the contrary \cite{liesche2013a}, be sufficient to load the sucrose through narrow plasmodesmata into the phloem of a polymer trap plant, while efficiently blocking the back flow of larger sugars. The simultaneous water flow can also be of importance not only to support the sugar flux but also to achieve advantageous pressure relations in the leaf and thus to preserve the vital functions of the strongly interdependent phloem and xylem vascular systems. We show that the bulk water entering the symplasm of pre-phloem cells already outside the veins can effectively suffice to drive the M{\"u}nch flow, although the same flow does only contribute a minor part to the loading of sugar into the intermediary cells of the phloem.
\section{The polymer trap model}
\begin{figure}[htbp]
\centering
\includegraphics[width=\textwidth]{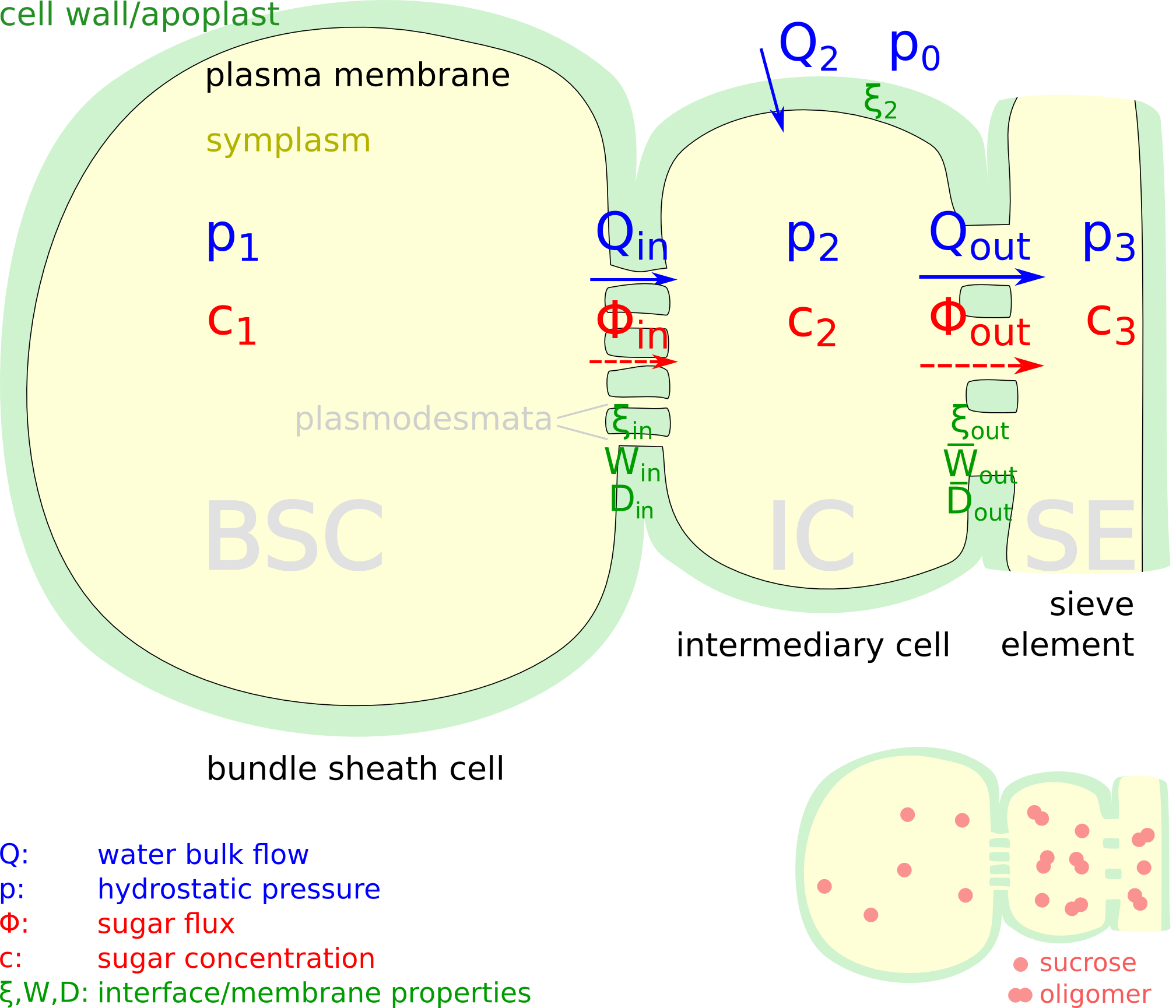}
\caption{(color online) \textbf{The polymer trap model with diffusion and bulk flow.} The water flow rates $Q$ through the cell interfaces and IC membrane are depicted with blue (full) arrows, the sugar flow rates $\Phi$ as red (dashed) arrows. These flows depend on the pressures $p$ as well as on sucrose and oligomer concentrations $c$ inside and outside the cells on the loading pathway. The semi-permeable cell interfaces are characterized by the permeability $\xi$, the bulk hindrance factor $W$, and the effective diffusion coefficient $D$ with subscripts `in' and `out'. Bundle sheath cell (BSC), intermediary cell (IC), and sieve element (SE) are numbered according to the loading steps. The IC and SE are both part of the phloem and are well connected via wide contacts called {\em pore plasmodesma units}. The BSC-IC interface is characterized by narrow plasmodesmata (PDs), which prevent the oligomers from diffusing back into the bundle sheath.}
\label{polymertrap}
\end{figure}
The polymer trap loading mechanism was postulated for angiosperm taxa, for example, cucurbits, and is shown in Fig. \ref{polymertrap}.\\
Most of the concrete values which are used in our calculations, i.e., the sugar concentrations in the cells of the loading pathway \cite{haritatos1996a}, the surface and interface areas of the cells \cite{volk1996a}, and the total leaf sugar export \cite{schmitz1987a}, were measured in muskmelon (\textit{Cucumis melo}). The cytosolic concentration of sucrose is around $\unit[200]{mM}$ \cite{haritatos1996a} in the mesophyll and bundle sheath cells (BSCs) taking into account the intracellular compartmentation. Sucrose passes symplasmically through narrow plasmodesmata (PDs) into the companion cells of the phloem, which are called {\em intermediary cells} (ICs) in this special loading type. In the ICs the sucrose is converted to larger oligomers, also called raffinose family oligosaccharides (RFOs), which pass through relatively wide PDs into the sieve element (SE). The tetrasaccharide stachyose is the most abundant sugar oligomer in the phloem of \textit{Cucumis melo}. The sucrose and stachyose concentrations in the phloem {\em cytosol}, i.e., in the cell sap outside of the vacuole, were measured to be about $\unit[132]{mM}$ and $\unit[335]{mM}$, respectively \cite{haritatos1996a}.  These two sugars represent together about $87 \%$ of the total sugar concentration in the phloem, which, with a value of $\unit[539]{mM}$, is more than twice as large as the concentration in the bundle sheath cytosol \cite{haritatos1996a}. \\

On the contrary, almost no RFOs have been found outside the SE-IC complex, and since no evidence for active sucrose transporters in the bundle sheath membranes of RFO-transporting plants have been found, it seems that the narrow plasmodesmatal pores in the BSC-IC interface must provide the delicate filtering effect  letting the smaller sucrose molecules pass from the bundle sheath while retaining the oligomers in the phloem \cite{rennie2009a}.
For this task, the effective pore widths must be similar to the diameters of the sugar molecules i.e., around $\unit[1]{nm}$. Such small widths seem at least not in conflict with evidence from electron microscopy, where parts of the plasmodesmata found in the IC wall look totally obstructed  \cite{fisher1986a}, but where one can hardly resolve patterns of sizes below $\unit[1]{nm}$. Schmitz \textit{et al.} measured the total export rate in leaves of \textit{Cucumis melo} \cite{schmitz1987a}, from which a sugar current density $J_\mathrm{in}\approx \unit[9.7\cdot 10^{-7}]{mol\,m^{-2}\,s^{-1}}$ across the BSC-IC interface can be calculated \cite{liesche2013a}.\\

The explanation of the functioning of the polymer trap given by Turgeon and collaborators \cite{rennie2009a} is that the sucrose diffuses along a downhill concentration gradient into the phloem while the oligomers, which are synthesized by enzymatic reactions at this location, are blocked by the specialized narrow PDs in the IC wall from diffusing back into the bundle sheath. This simple picture was questioned by Liesche and Schulz \cite{liesche2013a}, who considered quantitatively the hindered diffusion across the BSC-IC interface. 
In the present paper, we present an extended model, relating the transport coefficients to the structure and density of PDs in the cellular interfaces, and including explicitly the water flow. Based on the available experimental data, we show that pure diffusion {\em can} create a large enough sugar export in \textit{Cucumis melo} while blocking the oligosaccharides, but since the pores are of the dimension of the sugar molecules, osmotic effects across the cell interfaces are unavoidable and probably important. Thus, the resulting water flows may be crucial for building up the bulk flow in the phloem vascular system. We calculate the hydrostatic pressures created in the cells, and to compute a possible water intake across the cell membranes, we have to compare the resulting water potentials to that of the apoplast outside the cell membranes. We expect the pressures in the apoplast to be close to the (negative) values in the xylem, which are unfortunately not known for this particular species. However, we assume the value in musk melon to be close to that in maize, which has a typical xylem pressure of around $\unit[-4]{bar}$ \cite{tyree2002a}. The (positive) so-called turgor pressure for well-hydrated living cells should be large enough to keep the fairly elastic plasma membrane tight against the rigid cell wall. Since there are, as far as we know, no data available for the leaf cell pressures in \textit{Cucumis melo} we assume them to be larger than and close to the ambient pressure similar to the mesophyll turgor pressures measured in \textit{Tradescantia virginiana} \cite{zimmermann1980a}. We use the lower limit $\unit[1]{bar}$ as a reasonable value for the bundle sheath pressure in our numerical calculations.  With this assumption the pressure in the phloem thus builds up to values of close to 10 bars, which is a typical value quoted for the phloem pressure \cite{Nobel, Taiz}.

\subsection{Transport equations for the polymer trap model}

Our model (see Fig. \ref{polymertrap}) considers diffusion and bulk flow through the plasmodesmata of the  BSC-IC and IC-SE cell interfaces and furthermore takes into account a possible osmotic water flow across the IC-plasma membrane. For simplicity we assume here that, in the IC, two sucrose molecules are oligomerized to one tetrasaccharide, corresponding to a stachyose molecule in \textit{Cucumis melo}. The volume and sugar flows across the two cell interfaces can be written using the Kedem-Katchalsky equations \cite{kedem1958a} for membrane flows in the presence of multiple components. The volumetric water flow rates (measured, e.g., in $\unit{m^3\, s^{-1}}$) into and out of the IC can be expressed as
\begin{align}
Q_\mathrm{in}&=\xi_\mathrm{in}\left[(c_2^\mathrm{s}-c_1^\mathrm{s})(1-W_\mathrm{in}^\mathrm{s})RT+c_2^\mathrm{o}RT-(p_2-p_1)\right]
\label{QIn}\\
&=\xi_\mathrm{in}\left[\Psi_1-\Psi_2+W_\mathrm{in}^\mathrm{s}\Delta c_\mathrm{in}^\mathrm{s} R T  \right]\nonumber \\
Q_\mathrm{out}&=\xi_\mathrm{out}\left[(c_3^\mathrm{s}-c_2^\mathrm{s})(1-W_\mathrm{out}^\mathrm{s})RT+(c_3^\mathrm{o}-c_2^\mathrm{o})(1-W_\mathrm{out}^\mathrm{o})RT-(p_3-p_2)\right]
\label{QOut}\\
&=\xi_\mathrm{out}\left[\Psi_2-\Psi_3+W_\mathrm{out}^\mathrm{s}(c_2^\mathrm{s}-c_3^\mathrm{s})RT+W_\mathrm{out}^\mathrm{o}(c_2^\mathrm{o}-c_3^\mathrm{o})RT\right].\nonumber
\end{align}
where the subscripts number the cells in the sequence BSC, IC, SE, and $\Delta c_\mathrm{in}^\mathrm{s}=c_1^\mathrm{s}-c_2^\mathrm{s}$. The superscripts denote the molecule species, sucrose (s) and oligomer (o). The water potentials are defined as $\Psi_i = p_i-RTc_i$. Note that the water can flow through the plasmodesmata from a lower to a higher water potential because of the different osmotic effects of the sugar species. 
The coefficients $W$ are the bulk hindrance factors $W = 1-\sigma$, where $\sigma$ is the reflection coefficient used by Kedem and Katchalsky. Thus, if $W=0$ for a given molecule, it cannot get through the membrane and creates a full osmotic pressure, while $W=1$ means that the molecule passes as easily as the water molecules. We use the universal gas constant $R=\unit[8.314]{J\, mol^{-1}\, K^{-1}}$ and the absolute temperature $T=\unit[300]{K}$.

The corresponding sugar flow rates (e.g., in $\unit{mol\, s^{-1}}$) can then be written as
\begin{align}
\Phi_\mathrm{in}&=Q_\mathrm{in}c_1^\mathrm{s}W_\mathrm{in}^\mathrm{s}+\frac{A_\mathrm{in}}{d}D_\mathrm{in}^\mathrm{s}\Delta c_\mathrm{in}^\mathrm{s}
\label{PhiIn}\\
\Phi_\mathrm{out}&=Q_\mathrm{out}\left[c_2^\mathrm{s}W_\mathrm{out}^\mathrm{s}+c_2^\mathrm{o}W_\mathrm{out}^\mathrm{o}\right]+\frac{A_\mathrm{out}}{d}\left[D_\mathrm{out}^\mathrm{s}(c_2^\mathrm{s}-c_3^\mathrm{s})+D_\mathrm{out}^\mathrm{o}(c_2^\mathrm{o}-c_3^\mathrm{o})\right]
\label{PhiOut}
\end{align}
Here $D$ is a diffusion coefficient related to the diffusive mobility $\omega$ used by Kedem and Katchalsky as $D=d\omega R T$. $A$ is an interfacial area and $d$ is the diffusion distance, i.e., the thickness of the intermediary cell wall.  The two terms in $\Phi$ describe, respectively, the advective contribution (proportional to $Q$) and the diffusive one (proportional to the concentration differences). The interface coefficients are computed in the next section, based upon the geometry of the PDs. 

If we introduce average interface coefficients $\bar{W}_\mathrm{out}=(x^\mathrm{s}W_\mathrm{out}^\mathrm{s}+x^\mathrm{o}W_\mathrm{out}^\mathrm{o})$ and $\bar{D}_\mathrm{out}=(x^\mathrm{s}D_\mathrm{out}^\mathrm{s}+x^\mathrm{o}D_\mathrm{out}^\mathrm{o})$ with the sucrose and oligomer proportions $x^\mathrm{s(o)}  =c_2^\mathrm{s(o)}/c_2=c_3^\mathrm{s(o)}/c_3$ in the phloem, the expressions (\ref{QOut}) and (\ref{PhiOut}) for the outflows can be simplified to
\begin{align}
Q_\mathrm{out}&=\xi_\mathrm{out}\left[(c_3-c_2)(1-\bar{W}_\mathrm{out})RT-(p_3-p_2)\right]
\label{QOutSimp}\\
&=\xi_\mathrm{out}\left[\Psi_2-\Psi_3+(c_2-c_3)\bar{W}_\mathrm{out} R T\right]\nonumber \\
\Phi_\mathrm{out}&=Q_\mathrm{out}c_2 \bar{W}_\mathrm{out}+\frac{A_\mathrm{out}}{d}\bar{D}_\mathrm{out}(c_2-c_3),
\label{PhiOutSimp}
\end{align}
where we assume that the sucrose and oligomer proportions are the same in the SE and the IC.
There might also be an osmotic water flow $Q_2$ across the IC membrane, which builds a connection to the apoplast, where we expect a (negative) hydrostatic pressure $p_0$, probably close to the xylem pressure. This trans-membrane flow can be written using the permeability coefficient $\xi_2$ and the van't Hoff equation for an ideally semi-permeable IC membrane as
\begin{align}
Q_2=\xi_2 \left[R T c_2-(p_2-p_0)\right]=\xi_2 \left[p_0-\Psi_2\right].
\label{Q2}
\end{align} 
For a water flow $Q_2>0$ into the intermediary cell the water potential $\Psi_2=p_2-R T c_2$ has to be less (more negative) than the pressure $p_0$ in the apoplast.
The flows into and out of the IC are related by conservation laws for water and sugar in the form
\begin{align}
Q_\mathrm{in} &+Q_2=Q_\mathrm{out}
\label{eqVolCons}\\
 \Phi_\mathrm{in}&=(x^\mathrm{s}+2x^\mathrm{o})\Phi_\mathrm{out},
\label{eqMassCons}
\end{align}
where Eq. (\ref{eqMassCons}) is derived from the mass conservation $M^\mathrm{s}\Phi_\mathrm{in}=\frac{1}{c_2}(M^\mathrm{s}c_2^\mathrm{s}+M^\mathrm{o}c_2^\mathrm{o})\Phi_\mathrm{out}$ of sugar molecules in the intermediary cell with the molar masses related by $M^\mathrm{o}=2 M^\mathrm{s}$ used in our approximate model.

\section{Estimates of the coefficients and concentrations}
\label{secCoeff}
\begin{figure}[ht]
\centering
\includegraphics[width=\textwidth]{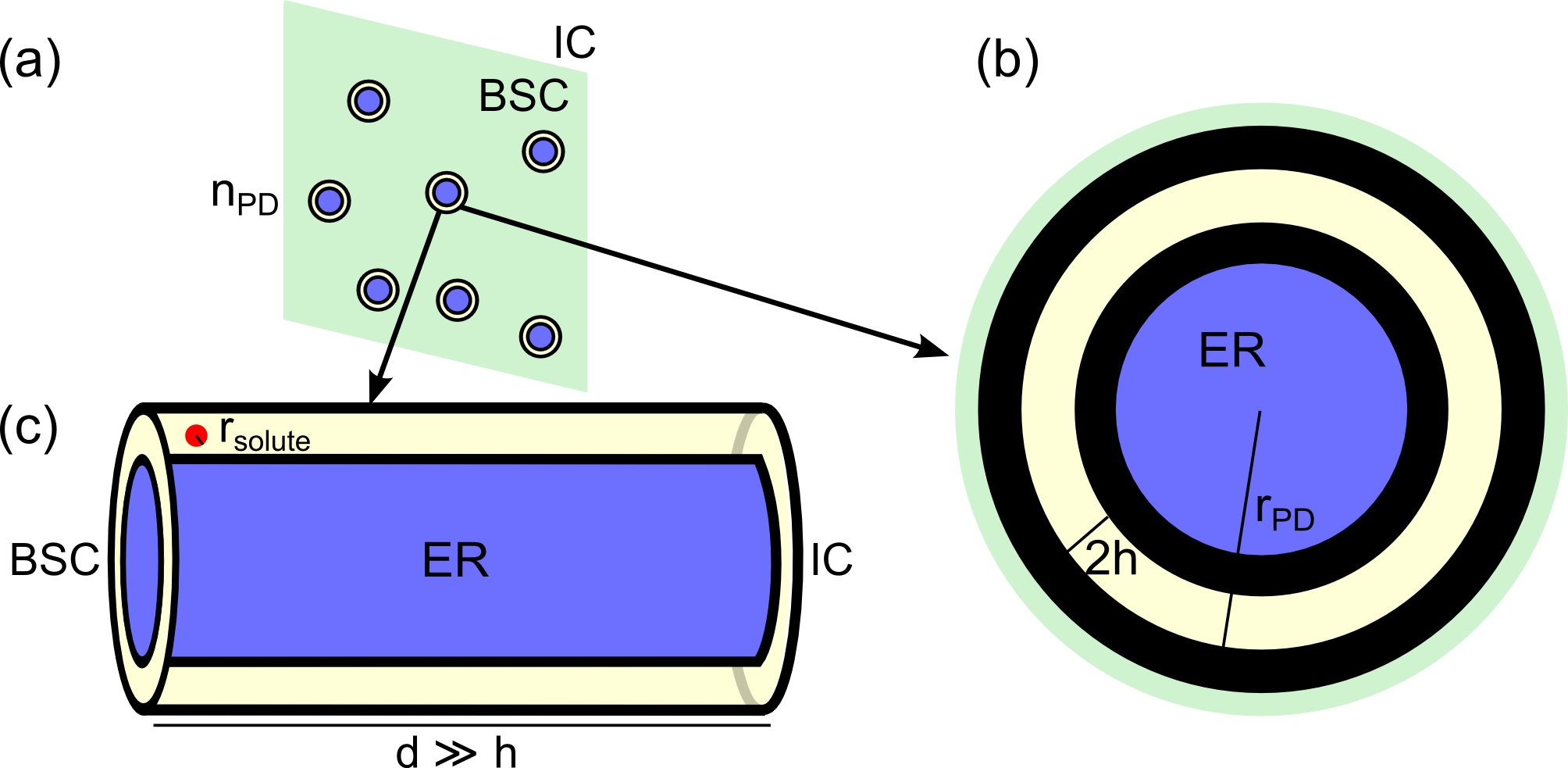}
\caption{(color online) \textbf{Three perspectives of the plasmodesmata modeled as slit pores.}
Part of the cell wall between BSC and IC with PD density $n_\mathrm{PD}$ is sketched in (a). The assumed substructure of a PD is shown in cross section (b) and three dimensionally (c). The cytoplasmic sleeve (light yellow) available for water and sugar transport is restricted by the desmotubule of the endoplasmic reticulum [ER, blue (gray)] and electron-dense particles (black) attached to the membrane, and is assumed to take the form of a circular slit with radius $r_\mathrm{PD}$, half-width $h$, and length $d$.}
\label{figSlitPerspectives}
\end{figure}
\begin{table}
\centering
\begin{tabular}{l p{0.6\textwidth} l l l}
\hline
Variable & Measured as & Value & Unit & Reference\\
\hline
$A_\mathrm{in}$ & Interface area between IC and BSC & $10^{-9}$ & $\unit{m^2}$ & \cite{volk1996a}\\
$A_\mathrm{out}$ & Interface area between IC and SE & $0.2\cdot 10^{-9}$ & $\unit{m^2}$ & \cite{volk1996a}\\
$A_2$ & Surface area of the IC & $10^{-9}$ & $\unit{m^2}$ & \cite{volk1996a}\\
$r^\mathrm{s}$ & Hydrodynamic radius of sucrose from 3D-model & $4.2\cdot 10^{-10}$ & $\unit{m}$ & \cite{liesche2013a}\\
$r^\mathrm{o}$ & Hydrodynamic radius of stachyose from 3D-model & $6.0\cdot 10^{-10}$ & $\unit{m}$ & \cite{liesche2013a}\\
$D^\mathrm{s}=1/2\,D^\mathrm{s}_\mathrm{water}$ & Free cytosolic diffusion coefficient for sucrose & $2.3\cdot 10^{-10}$ & $\unit{m^2\, s^{-1}}$ & \cite{henrion1964a}\\
$D^\mathrm{o}=1/2\,D^\mathrm{o}_\mathrm{water}$ & Free cytosolic diffusion coefficient for stachyose & $1.9\cdot 10^{-10}$ & $\unit{m^2\, s^{-1}}$ & \cite{craig1962a}\\
$f^\m{s}$ & Shape factor for hydrated sucrose molecules & $0.88$ &\\
$f^\m{o}$ & Shape factor for hydrated stachyose molecules & $1.04$& \\
$\eta_\mathrm{cyt}$ & Dynamic viscosity of cytosol & $2\cdot 10^{-3}$ & $\unit{Pa\, s}$ & \cite{liesche2013a}\\
$h_\mathrm{in}$ & Half-slit width of PDs in the IC wall & $<10^{-9}$ & $\unit{m}$ & \cite{fisher1986a,roberts2003a}\\
$h_\mathrm{out}$ & Half-slit width of "normal" PDs & $10^{-9}$ & $\unit{m}$ & \cite{roberts2003a}\\
$r_\mathrm{PD}$ & Average radius of PDs in plant cell walls & $2.5\cdot 10^{-8}$ & $\unit{m}$ & \cite{fisher1986a,roberts2003a}\\
$d$ & Thickness of the IC wall & $10^{-7}$ & $\unit{m}$ & \cite{volk1996a}\\
$n_\mathrm{PD}$ & Density of PDs in the IC wall & $10^{13}$ & $\unit{m^{-2}}$ & \cite{volk1996a}\\
$c_1=c_1^\mathrm{s}$ & Cytosolic sucrose concentration in mesophyll and bundle sheath & $200$ & $\unit{mol\, m^{-3}}$ & \cite{haritatos1996a}\\
$c_2$ & Total cytosolic sugar concentration in the IC-SE complex & $500$ & $\unit{mol\, m^{-3}}$ & \cite{haritatos1996a}\\
$c_2^\mathrm{s}$ & Cytosolic sucrose concentration in IC-SE complex & $140$ & $\unit{mol\, m^{-3}}$ & \cite{haritatos1996a}\\
$\Delta c_\m{in}^\mathrm{s}=c_1^\m{s}-c_2^\m{s}$ & Sucrose concentration difference between BSC- and IC-cytosol & $60$ & $\unit{mol\, m^{-3}}$ & \cite{liesche2013a, haritatos1996a}\\
$p_1$ & Hydrostatic pressure in the bundle sheath & $\sim 10^5$ & $\unit{Pa}$ & \cite{zimmermann1980a}\\
$p_0$ & Xylem and apoplast pressure (from maize) & $ -4\cdot 10^{5}$ & $\unit{Pa}$ & \cite{tyree2002a}\\
$J_\mathrm{in}=\Phi_\m{in}/A_\m{in}$ & Sugar current density through BSC-IC interface, from total leaf export rate & $9.7\cdot 10^{-7}$& $\unit{mol\,m^{-2}\,s^{-1}}$ & \cite{schmitz1987a} \\
\hline
\end{tabular}
\caption{Parameter values characterizing the loading pathway in \textit{Cucumis melo},\\ estimated from the given references.}
\label{tabValues}
\end{table}
The cell interfaces are modeled as porous membranes. From detailed electron microscopic investigations  \cite{fisher1986a,volk1996a} the PDs at this specific interface  are generally branched towards the IC. However, the detailed substructure is not known, in particular the shape and area of the {\em cytoplasmic sleeve} connecting the cytosol of the cells.
For our modeling we  simplify these channels as circular slits (see Fig. \ref{figSlitPerspectives}), as suggested in Ref. \cite{waigmann1997a}, with average radius $r_\mathrm{PD}$, half-width $h\leq \unit[1]{nm}$, and length $d$ equal to the thickness of the part of the cell wall belonging to the IC.\\

\begin{figure}[ht]
\centering
\includegraphics{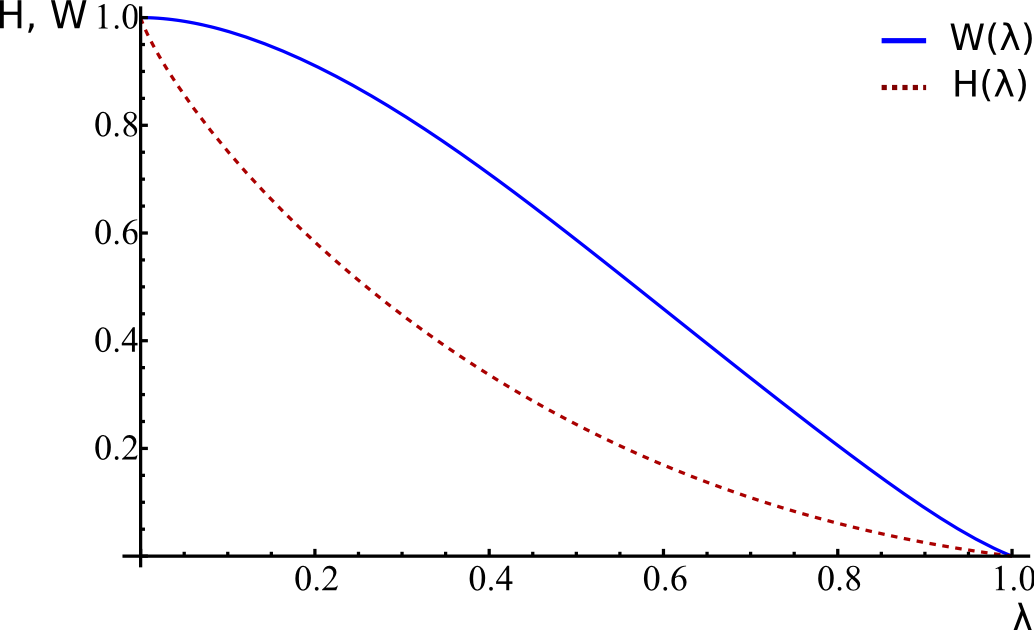}
\caption{(color online) \textbf{Diffusive and convective hindrance factors $H$ (blue, solid) and $W$ (red, dashed) in circular slit pores as function of the relative solute size $\lambda$.} Both approximations given by Ref. \cite{dechadilok2006a} decrease smoothly from 1 to 0 for an increasing solute size, where a hindrance factor of zero corresponds to total blockage of the respective molecule. The convective hindrance factor $W$ is in the whole range larger than the diffusive hindrance factor $H$. Above $\lambda=0.8$ the curves should be regarded as extrapolations.}
\label{hindrancePlots}
\end{figure}

From the slit geometry together with the density $n_\mathrm{PD}$ of plasmodesmata and the interface areas $A_\mathrm{in(out)}$ (see Table \ref{tabValues}) the interface coefficients can be calculated using the hindrance factors $H$ and $W$ for diffusion and convection in narrow pores, which were recently analyzed by Deen and Dechadilok \cite{dechadilok2006a}. For spherical particles these hindrance factors have been estimated as polynomials in the relative solute size $\lambda=r_\mathrm{solute}/h$. The following expressions are valid for $0\leq \lambda\leq 0.8$ (H) and $0\leq \lambda \leq 0.95$ (W),
\begin{align}
H(\lambda) 
&=1+\frac{9}{16}\lambda\ln\lambda-1.19358\lambda+0.4285\lambda^3-0.3192\lambda^4+0.08428\lambda^5
\label{defH}\\
W(\lambda)&=1-3.02\lambda^2+5.776\lambda^3-12.3675\lambda^4+18.9775\lambda^5-15.2185\lambda^6+4.8525\lambda^7.
\label{defW}
\end{align}
For $\lambda\geq 1$ the solute should be totally blocked by the plasmodesmatal pores. In this case both hindrance factors are set to zero. Plots of the hindrance factors as functions of $\lambda$ are shown in Fig. \ref{hindrancePlots}.\\
The bulk hindrance factor $W_\mathrm{in(out)}^\mathrm{s(o)}$ enters our equations directly as one of the three interface coefficients.
The diffusive hindrance factor $H_\mathrm{in(out)}^\mathrm{s(o)}$ is used together with the pore covering fraction $\gamma_\m{in(out)}$ to compute the effective diffusion coefficients $D_\mathrm{in(out)}^\mathrm{s(o)}$ appearing in (\ref{PhiIn}) and (\ref{PhiOut}) as
\begin{align}
D_\mathrm{in(out)}^\mathrm{s(o)}=\gamma_\mathrm{in(out)}H_\mathrm{in(out)}^\mathrm{s(o)}D^\mathrm{s(o)}.
\label{effDiffCoeff}
\end{align}
Here the covering fraction $\gamma_\m{in(out)}$ is given as the ratio of free slit-space to total cell-interface area, i.e.,
\begin{align}
\gamma_\mathrm{in(out)}=4\pi r_\mathrm{PD}h_\mathrm{in(out)}\cdot n_\mathrm{PD},
\label{coveringFrac}
\end{align}
where $n_\m{PD}$ is the density of plasmodesmata in the IC wall, and the unobstructed sleeve is assumed to be very narrow ($h_\m{in(out)}\ll r_\m{PD}$). 
The free diffusion coefficient $D^\m{s(o)}$ of the respective solutes in cytosol can be written using the Einstein relation for diffusing spherical molecules as 
\begin{align}
D^\m{s(o)}=\frac{kT}{6\pi  \eta_\m{cyt} r^\m{s(o)}}f^\m{s(o)}
\label{DEinstein}
\end{align}
with the hydrodynamic radii $r^\m{s(o)}$ of the solutes, the cytosolic viscosity $\eta_\m{cyt}$ and the Boltzmann constant $k$ related to the universal gas constant $R=N_\m{A} k$ by the Avogadro constant $N_\m{A}=\unit[6\cdot 10^{23}]{mol^{-1}}$.
The shape factor $f$ accounts for the deviation from the Einstein relation primarily due to the non spherical shape of the molecule. In our modeling we use a three-dimensional (3D) structural model to compute the radii $r^\m{s(o)}$ for hydrated molecules \cite{liesche2013a} and thus include shape factors of the order of unity (see table \ref{tabValues}).\\
The permeability coefficient $\xi_\mathrm{in(out)}$ for the BSC-IC and IC-SE-interface is estimated using a pressure driven Poiseuille-flow $Q_\mathrm{slit}$ through narrow rectangular channels of height $2h$ and width $2\pi r_\mathrm{PD}$, where $h_\mathrm{in(out)}\ll r_\mathrm{PD}$, i.e.
\begin{align}
A_\mathrm{in(out)}n_\mathrm{PD}Q_\mathrm{slit}&=A_\mathrm{in(out)}n_\mathrm{PD}\frac{4 \pi r_\mathrm{PD}h_\mathrm{in(out)}^3}{3\eta_\mathrm{cyt}d}\Delta p=\xi_\mathrm{in(out)}\Delta p\\
\Rightarrow \xi_\mathrm{in(out)}&=A_\mathrm{in(out)}n_\mathrm{PD}\frac{4 \pi r_\mathrm{PD}h_\mathrm{in(out)}^3}{3\eta_\mathrm{cyt}d}.
\label{permeability}
\end{align} 
The cytosolic viscosity is estimated with a value twice as large as the viscosity of water, i.e., $\eta_\mathrm{cyt}=\unit[2\cdot 10^{-3}]{Pa\, s}$. The characteristic cell-wall thickness $d$ as well as the plasmodesmata radius $r_\mathrm{PD}$ have been estimated from TEM-images \cite{volk1996a, botha1993a}. Based on the measurements by Volk \textit{et al.}, the density $n_\mathrm{PD}$ of plasmodesmata in the IC wall is fixed to a value of around $\unit[10]{\mu m^{-2}}$ \cite{volk1996a}. For the BSC-IC interface we assume that the PDs are very narrow and have a half-width between the hydrodynamic radius of sucrose $r^\mathrm{s}\approx \unit[0.42]{nm}$ and of stachyose $r^\mathrm{o}\approx\unit[0.60]{nm}$, since stachyose should be totally blocked from going back to the bundle sheath. We shall choose $h_\mathrm{in}=r^\mathrm{o}=\unit[0.6]{nm}$ as a standard value since it is the largest value for which we are certain that $W_\mathrm{in}^\mathrm{o}=H_\mathrm{in}^\mathrm{o}=0$ (see, however, the final section on raffinose hindrance). The hydrodynamic radii $r^\mathrm{s}$ and $r^\mathrm{o}$ have been computed using the 3D-structural models of hydrated sucrose and stachyose molecules accounting in particular for the cylindrical molecule forms \cite{liesche2013a}. For the IC-SE interface, the PDs are wider and we use a ``normal" slit-width
$h_\mathrm{out}=\unit[1]{nm}$ \cite{roberts2003a}. The interface coefficients for this configuration are listed in table \ref{tabCoefficients}.

The sucrose and total sugar concentrations in the IC are fixed to the values $\unit[140]{mM}$ and $\unit[500]{mM}$, respectively (see Table \ref{tabValues}), based on the measured concentrations from Ref. \cite{haritatos1996a}.

\begin{table}
\centering
\begin{tabular}{l l r}
\hline
Coefficient & Value & Unit\\
\hline
$W_\mathrm{in}^\mathrm{s}$ & $0.33$&\\
$W_\mathrm{out}^\mathrm{s}$ & $0.69$&\\
$W_\mathrm{out}^\mathrm{o}$ & $0.46$&\\
$D_\mathrm{in}^\mathrm{s}$ & $4.71\cdot 10^{-14}$ & $\unit{m^2\,s^{-1}}$\\
$D_\mathrm{out}^\mathrm{s}$ & $2.29\cdot 10^{-13}$ & $\unit{m^2\,s^{-1}}$\\
$D_\mathrm{out}^\mathrm{o}$ & $1.01\cdot 10^{-13}$ & $\unit{m^2\,s^{-1}}$\\
$\xi_\mathrm{in}$ & $1.13\cdot 10^{-21}$ & $\unit{m^3\,Pa^{-1}\,s^{-1}}$\\
$\xi_\mathrm{out}$ & $1.05\cdot 10^{-21}$ &$\unit{m^3\,Pa^{-1}\,s^{-1}}$\\
\hline
\end{tabular}
\caption{Calculated interface coefficients for the half-slit widths $h_\mathrm{in}=\unit[0.6]{nm}$ and $h_\mathrm{out}=\unit[1]{nm}$.}
\label{tabCoefficients}
\end{table}

\section{Dimensionless equations and their solution}

\begin{table}[ht]
\centering
\begin{tabular}{l l l}
\hline
Variable & Scaling factor & Value\\ \hline
$A$ & $A_\mathrm{in}$ & $\unit[10^{-9}]{m^2}$\\
$c$ & $c_1$ & $\unit[200]{mol\,m^{-3}}$ ($\unit[200]{mM}$)\\
$p$ & $R T c_1$ & $\unit[5\cdot 10^{5}]{Pa}$ ($\unit[5]{bar}$)\\
$\Psi$ & $R T c_1$ & $\unit[5 \cdot 10^{5}]{Pa}$\\
$\xi$ & $\xi^*=\xi_\mathrm{in}(h_\mathrm{in}=r^\mathrm{s})$ & $\unit[4\cdot 10^{-22}]{m^3\,Pa^{-1}\,s^{-1}}$\\
$D$ & $R T d \xi^* c_1/A_\mathrm{in}$ & $\unit[2\cdot 10^{-14}]{m^2\, s^{-1}}$\\
$Q$ & $\xi^* R T c_1$ & $\unit[2\cdot 10^{-16}]{m^3\, s^{-1}}$\\
$\Phi$ & $\xi^* R T c_1^2$ & $\unit[4\cdot 10^{-14}]{mol\,s^{-1}}$\\
$J_\m{in}$ & $\xi^* R T c_1^2/A_\m{in}$ & $\unit[4\cdot 10^{-5}]{mol\,m^{-2}\,s^{-1}}$ \\
\hline
\end{tabular}
\caption{Scaling factors for the non-dimensionalization.}
\label{tabScaling}
\end{table}

To nondimensionalize we scale the used variables with the factors stated in Table \ref{tabScaling} based on the concentration $c_1$ in the BSC and the properties of the BSC-IC interface. The dimensionless flows can be written as
\begin{align}
\hat{Q}_\mathrm{in}&=\hat{\xi}_\mathrm{in}\left[\hat{c}_2^\mathrm{o}-(1-W_\mathrm{in}^\mathrm{s})\Delta \hat{c}_\mathrm{in}^\mathrm{s}-(\hat{p}_2-\hat{p}_1)\right]\nonumber \\
&=\hat{\xi}_\mathrm{in}\left[\hat{\Psi}_1-\hat{\Psi}_2+W_\mathrm{in}^\mathrm{s}\Delta \hat{c}_\mathrm{in}^\mathrm{s}\right]
\label{QInDimless}\\
\hat{Q}_\mathrm{out}&=\hat{\xi}_\mathrm{out}\left[(1-\bar{W}_\mathrm{out})(\hat{c}_3-\hat{c}_2)-(\hat{p}_3-\hat{p}_2)\right]\nonumber \\
&=\hat{\xi}_\mathrm{out}\left[\hat{\Psi}_2-\hat{\Psi}_3+\bar{W}_\mathrm{out}(\hat{c}_2-\hat{c}_3)\right]
\label{QOutDimless}\\
\hat{Q}_2&=\hat{\xi}_2\left[\hat{c}_2-(\hat{p}_2-\hat{p}_0)\right]\nonumber \\
&=\hat{\xi}_2\left[\hat{p}_0-\hat{\Psi}_2\right]
\label{Q2Dimless}\\
\hat{\Phi}_\mathrm{in}&=W_\mathrm{in}^\mathrm{s}\hat{Q}_\mathrm{in}+\hat{D}_\mathrm{in}^\mathrm{s}\Delta \hat{c}_\mathrm{in}^\mathrm{s}
\label{PhiInDimless}\\
\hat{\Phi}_\mathrm{out}&=\bar{W}_\mathrm{out}\hat{Q}_\mathrm{out}\hat{c}_2 +\hat{A}_\mathrm{out}\hat{\bar{D}}_\mathrm{out}(\hat{c}_2-\hat{c}_3)
\label{PhiOutDimless}
\end{align}

In addition we have the conservation laws (\ref{eqVolCons}) and (\ref{eqMassCons}), which are unchanged, i.e.,
\begin{align}
\hat{Q}_\mathrm{in} &+\hat{Q}_2=\hat{Q}_\mathrm{out}
\label{eqVolConsDimless}\\
\hat{ \Phi}_\mathrm{in}&=(x^\mathrm{s}+2x^\mathrm{o})\hat{\Phi}_\mathrm{out}.
\label{eqMassConsDimless}
\end{align}

The dimensionless sugar inflow corresponding to the experimentally determined sugar current density $J_\m{in}=\unit[9.7\cdot 10^{-7}]{mol\, m^{-2}\, s^{-1}}$ \cite{schmitz1987a} in \textit{Cucumis melo} is
\begin{align}
\hat{\Phi}_\m{in}^\m{exp}=\hat{J}_\m{in}=\frac{J_\m{in}A_\m{in}}{\xi^* R T c_1^2}=0.025.
\label{JHat}
\end{align} 
The scaled permeability $\hat{\xi}_\m{in(out)}$ and effective diffusion coefficients $\hat{D}_\m{in(out)}^\m{s(o)}$ take the form
\begin{align}
\hat{D}_\m{in(out)}^\m{s(o)}&=\frac{H_\m{in(out)}^\m{s(o)}f^\m{s(o)}}{(\lambda_\m{in(out)}^\m{s})^3 N_\m{in(out)}^\m{s(o)}}
\label{DHat}\\
\hat{\xi}_\m{in}&=(\lambda_\m{in}^\m{s})^{-3}
\label{xiInHat}\\
\hat{\xi}_\m{out}&=\hat{A}_\m{out}(\lambda_\m{out}^\m{s})^{-3}
\label{xiOutHat}
\end{align}
Here the definitions from Sec. \ref{secCoeff} and the scaling factors from Table \ref{tabScaling} were used, and the relative solute size in the slits of half-width $h_\m{in(out)}$ is defined as $\lambda_\m{in(out)}^\m{s(o)}=r^\m{s(o)}/h_\m{in(out)}$. The expression
$N_\m{in(out)}^\m{s(o)}=N_\m{A}c_1 2\pi (r^\m{s(o)})^3 (\lambda_\m{in(out)}^\m{s(o)})^{-2}$ can be understood as the average number of sucrose molecules in the BSC in a small volume $2\pi (r^\m{s(o)})^3 (\lambda_\m{in(out)}^\m{s(o)})^{-2}$ of the dimension of the sugar molecules.
Inserting the dimensionless coefficients in the scaled flows, these can be rewritten as
\begin{align}
\hat{Q}_\m{in}=&(\lambda_\m{in}^\m{s})^{-3}\left[\hat{\Psi}_1-\hat{\Psi}_2+W_\m{in}^\m{s}\Delta \hat{c}_\m{in}^\m{s}\right]
\label{QInHatRewritten}\\
\hat{Q}_\m{out}=&(\lambda_\m{out}^\m{s})^{-3}\hat{A}_\m{out}\left[\hat{\Psi}_2-\hat{\Psi}_3+(\hat{c}_2-\hat{c}_3)\bar{W}_\m{out}\right]
\label{QOutHatRewritten}\\
\hat{\Phi}_\m{in}=& W_\m{in}^\m{s}\hat{Q}_\m{in}+(\lambda_\m{in}^\m{s})^{-3}(N_\m{in}^\m{s})^{-1}H_\m{in}^\m{s}f^\m{s}\Delta \hat{c}_\m{in}^\m{s} 
\label{PhiInHatwithQ} \\
=&(\lambda_\m{in}^\m{s})^{-3}W_\m{in}^\m{s}\left[\hat{\Psi}_1-\hat{\Psi}_2\right]\nonumber \\
&+(\lambda_\m{in}^\m{s})^{-3}\left((W_\m{in}^\m{s})^2+(N_\m{in}^\m{s})^{-1}H_\m{in}^\m{s}f^\m{s}\right) \Delta \hat{c}_\m{in}^\m{s}
\label{PhiInHatRewritten}\\
\hat{\Phi}_\m{out}=& \bar{W}_\m{out}\hat{Q}_\m{out}\hat{c}_2+\hat{A}_\m{out}(\lambda_\m{out}^\m{s})^{-3}(\bar{N}_\m{out})^{-1}\bar{H}_\m{out}\bar{f} \left[\hat{c}_2-\hat{c}_3\right]
\label{PhiOutHatwithQ}\\
=&(\lambda_\m{out}^\m{s})^{-3}\hat{A}_\m{out}\bar{W}_\m{out}\left[\hat{\Psi}_2-\hat{\Psi}_3\right]\hat{c}_2\nonumber \\
&+(\lambda_\m{out}^\m{s})^{-3}\hat{A}_\m{out}\left(\bar{W}_\m{out}^2 \hat{c}_2+(\bar{N}_\m{out})^{-1}\bar{H}_\m{out}\bar{f}\right) \left[\hat{c}_2-\hat{c}_3\right].
\label{PhiOutHatRewritten}
\end{align}
The bar over a variable always denotes an average quantity, calculated with the proportions of the two different sugars in the phloem, e.g., $\bar{W}_\m{out}=x^\m{s}W_\m{out}^\m{s}+x^\m{o}W_\m{out}^\m{o}$ using the proportions $x^\m{s}=c_2^\m{s}/c_2$ and $x^\m{o}=1-x^\m{s}$ of sucrose and oligomer molecules in the phloem.\\

We can use, for example, $\Delta \hat{c}_\mathrm{in}^\mathrm{s}, x^\mathrm{o}, \hat{\Psi}_1, \hat{Q}_\mathrm{out}$ and $\hat{Q}_2$ as independent variables and calculate the other quantities. The sucrose and oligomer concentrations in the intermediary cell can be calculated from the concentration difference $\Delta \hat{c}_\mathrm{in}^\mathrm{s}$ between the BSC and the IC, and the oligomer proportion $x^\mathrm{o}$ in the phloem using e.g.,
$\hat{c}_2^{\m s} = 1 - \Delta \hat{c}_{\m in}^{\m s}$, $x^{\m s} = 1- x^{\m o}$, $\hat{c}_2 =\hat{c}_2^\m{s}/x^\m{s}$. 
The concentration $\hat{c}_3$ in the sieve element can then be determined from the volume and sugar conservation equations (\ref{eqVolConsDimless}) and (\ref{eqMassConsDimless}) with the use of expressions (\ref{PhiInHatwithQ}) and (\ref{PhiOutHatwithQ}) for the sugar flow rates, i.e.
\begin{align}
\hat{c}_3=\hat{c}_2&+\frac{(x^\mathrm{s}+2x^\mathrm{o})\hat{c}_2\bar{W}_\mathrm{out}-W_\mathrm{in}^\mathrm{s}}{(x^\mathrm{s}+2x^\mathrm{o})\hat{A}_\mathrm{out}\bar{H}_\m{out}\bar{f}}\cdot (\lambda_\m{out}^\m{s})^3\bar{N}_\m{out}\hat{Q}_\m{out}\nonumber \\
&+\frac{W_\mathrm{in}^\mathrm{s}}{(x^\mathrm{s}+2x^\mathrm{o})\hat{A}_\mathrm{out}\bar{H}_\m{out}\bar{f}}\cdot (\lambda_\m{out}^\m{s})^3\bar{N}_\m{out}\hat{Q}_2\nonumber \\
&-\frac{1}{(x^\mathrm{s}+2x^\mathrm{o})\hat{A}_\mathrm{out}}\cdot\frac{H_\m{in}^\m{s}f^\m{s}(\lambda_\m{out}^\m{s})^3\bar{N}_\m{out}}{\bar{H}_\m{out}\bar{f}(\lambda_\m{in}^\m{s})^3 N_\m{in}^\m{s}}\Delta \hat{c}_\mathrm{in}^\mathrm{s}.
\label{c3Sol}
\end{align}
Finally, using the expressions for the water flows (\ref{QInHatRewritten}), (\ref{QOutHatRewritten}), and (\ref{Q2Dimless}), the water potentials $\hat{\Psi}_2$, $\hat{\Psi}_3$, and $\hat{\Psi}_0$ and corresponding hydrostatic pressures $\hat{p}_i$ inside and outside the cells of the loading pathway can be calculated (with the interface coefficients from Table \ref{tabCoefficients} and the geometry as fixed in Table \ref{tabValues}) as
\begin{align}
\hat{\Psi}_2&=\hat{p}_2-\hat{c}_2=\hat{\Psi}_1-(\lambda_\m{in}^\m{s})^3(\hat{Q}_\mathrm{out}-\hat{Q}_2)+W_\mathrm{in}^\mathrm{s}\Delta \hat{c}_\m{in}^\m{s}
\label{Psi2Dimless}\\
\hat{\Psi}_3&=\hat{p}_3-\hat{c}_3=\hat{\Psi}_2-(\lambda_\m{out}^\m{s})^3\hat{A}_\m{out}^{-1}\hat{Q}_\mathrm{out}+\bar{W}_\mathrm{out}\left[\hat{c}_2-\hat{c}_3\right]
\label{Psi3Dimless}\\
\hat{\Psi}_0&=\hat{p}_0=\frac{\hat{Q}_2}{\hat{\xi}_2}+\hat{\Psi}_2.
\label{Psi0Dimless}
\end{align}

\section{Special cases}

\subsection{Pure Diffusion}
In this subsection we first investigate whether pure diffusion through plasmodesmata can transport enough sugar into the phloem, and, subsequently, whether this special case with no bulk flow through the  plasmodesmata represents a likely loading situation in real plants. Assuming that the sucrose is transported into the IC by pure diffusion without a supporting bulk flow, we get
\begin{align}
\hat{\Phi}_\mathrm{in}&=\hat{D}_\m{in}^\m{s}\Delta \hat{c}_\m{in}^\m{s}=\frac{H_\m{in}^\m{s}f^\m{s}}{N_\m{in}^\m{s}(\lambda_\m{in}^\m{s})^3}\Delta \hat{c}_\mathrm{in}^\mathrm{s}
\label{PhiInDiff}
\end{align} 
This is in agreement with Fick's first law of diffusion. Taking $r^\m{s}=\unit[0.42]{nm}$ gives $f^\m{s}=0.88$. The sugar current depends on the half-slit width $h_\mathrm{in}$ of the PDs in the BSC-IC interface through the relative solute size $\lambda_\m{in}^\m{s}$, which also appears as variable in the diffusive hindrance factor $H_\m{in}^\m{s}=H(\lambda=\lambda_\m{in}^\m{s})$.
\begin{figure}
\centering
\includegraphics{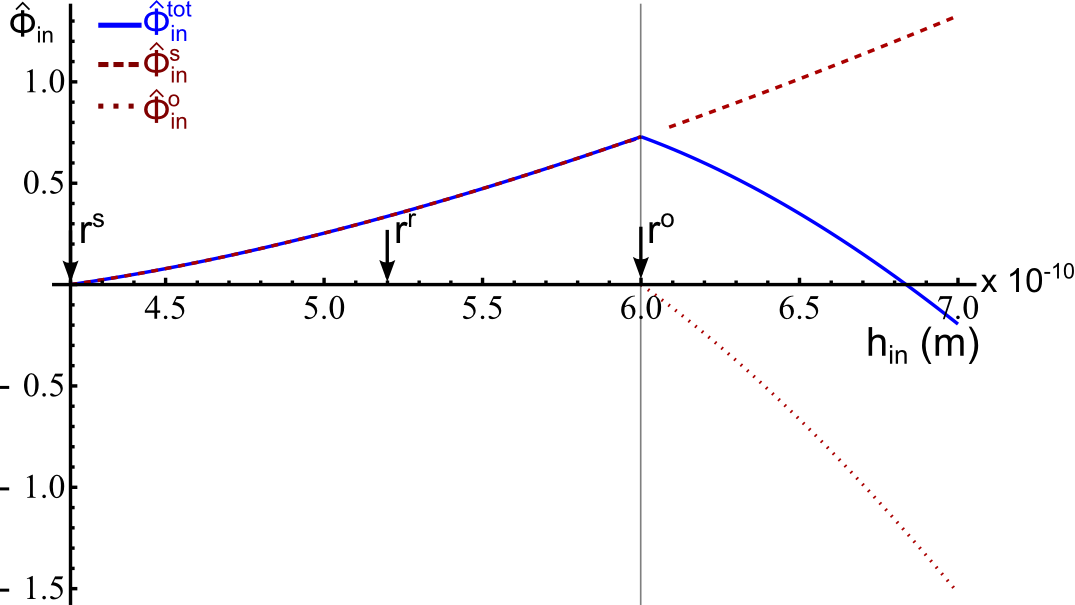}
\caption{(color online) \textbf{Sugar flow rate $\hat{\Phi}_\m{in}$ into the IC as function of the PD-half-slit width $h_\m{in}$ in the purely diffusive case.} The sugar flow rate is composed by the sucrose flow rate $\Phi_\m{in}^\m{s}$ (red, dashed) given by Eq. (\ref{PhiInDiff}) and the hypothetical negative oligomer flow rate $\Phi_\m{in}^\m{o}$ (red, dotted), which would occur when $h_\m{in}$ is larger than the oligomer radius $r^\m{o}$.
For the concentration differences measured in \textit{Cucumis melo} the flow rate $\hat{\Phi}_\m{in}^\m{o}$ of oligomers back into the bundle sheath would cause the total sugar flow rate $\hat{\Phi}_\m{in}^\m{tot}$ (blue, solid) to vanish at slit widths only about one tenth larger than these molecules. The diffusive sucrose flow rate $\hat{\Phi}_\m{in}^\m{s}$, however, gives a sufficient overall flux rate in the case of total blockage of the modeled oligomers (i.e., $h_\m{in}=r^\m{o}$) and even for smaller slits totally blocking raffinose molecules (i.e., $h_\m{in}=r^\m{r}$; see Sec. \ref{secraf}).}
\label{diffusionPlot}
\end{figure}
\\
Figure \ref{diffusionPlot} shows that even for slits which are only slightly larger than the oligomers, the back flow into the bundle sheath due to diffusion would exceed the sucrose flux in the opposite direction.
With our standard half-slit width of $h_\mathrm{in}=r^\mathrm{o}$ equal to the hydrodynamic radius of the stachyose molecules, corresponding to a relative sucrose size of $\lambda_\m{in}^\m{s}=0.7$, the tetrasaccharides in our model are blocked completely. For the sucrose flow rate we get $\hat{\Phi}_\m{in}=0.73$, which is about 30 times larger than the experimental value from Ref. \cite{schmitz1987a}. This shows that, in \textit{Cucumis melo}, diffusion through the narrow plasmodesmatal pores can be sufficient to achieve the measured sugar current into the phloem, and in fact the large value that we obtain probably means that the pores are even narrower than the size of the stachyose molecules. Indeed, the pores also have to be able to block the back flow of raffinose, which is around 10 \% smaller than stachyose. We discuss that in Sec. \ref{secraf} . \\

We found that pure diffusion is sufficient to export enough sugar into the phloem of RFO-transporting plants. On the other hand, the long-distance transport in the phloem system is based on a bulk flow for which water has to enter the symplasm at some point. Since in this special case we ruled out any bulk flow through the plasmodesmata between BSC and IC, the water has to go across the membrane of either the intermediary cell or the sieve element. We now calculate the pressures, concentrations, and water potentials in these cells to see if this is a possible and even advantageous situation for the plant, i.e., if the water potentials are low enough for water from the xylem to be drawn in.   
The condition of purely diffusive sugar loading implies that the hydrostatic and osmotic pressure differences across the BSC-IC interface must be balanced in order to achieve zero bulk flow. From this boundary condition, i.e., $\hat{Q}_\mathrm{in}=0$, the water potential $\hat{\Psi}_2$ and hydrostatic pressure $\hat{p}_2$ in the intermediary cell can be calculated for a fixed potential $\hat{\Psi}_1$ in the bundle sheath. With $\hat{Q}_\mathrm{in}=0$ Eq. (\ref{Psi2Dimless}) is reduced to
\begin{align}
\hat{\Psi}_2&=\hat{\Psi}_1+W_\m{in}^\m{s}\Delta \hat{c}_\m{in}^\m{s}
\label{p2ZeroQ}
\end{align} 
For a water potential of $\hat{\Psi}_1=-0.8$, corresponding to $p_1=\unit[1]{bar}$ in the bundle sheath, a value $\hat{\Psi}_2=-0.70$ results in the IC which corresponds to $\Psi_2=-\unit[3.5]{bar}$. To avoid inflow of water from the BSC, the intermediary cell thus has to build up a large hydrostatic pressure of $p_2=\unit[9.0]{bar}$. If the water needed in the phloem enters as $\hat{Q}_2>0$ across the membrane of the intermediary cell, the pressure in the apoplast has to be larger than the water potential $\hat{\Psi}_2$ in the IC, i.e., $p_0=Q_2/\xi_2+\Psi_2>\unit[-3.5]{bar}$. As mentioned above we assume the xylem pressure $p_0$ to be around $\unit[-4]{bar}$ \cite{tyree2002a}, and thus such a water uptake would not be feasible. For pressures \(p_1 > 1\) bar this conclusion is even more justified. \\
Now we consider the case $\hat{Q}_2=0$ where the flow through the PDs into the sieve element also vanishes, i.e., $\hat{Q}_\mathrm{out}=\hat{Q}_\mathrm{in}+\hat{Q}_2=0$. In this situation, the water from the xylem must flow in across the membrane of the sieve element. The concentration in the SE can be calculated with Eq. (\ref{c3Sol}), which simplifies for pure diffusion at both interfaces to 
\begin{align}
\hat{c}_3=\hat{c}_2-\frac{1}{(x^\mathrm{s}+2x^\mathrm{o})\hat{A}_\mathrm{out}}\cdot\frac{H_\m{in}^\m{s}f^\m{s}(\lambda_\m{out}^\m{s})^3\bar{N}_\m{out}}{\bar{H}_\m{out}\bar{f}(\lambda_\m{in}^\m{s})^3 N_\m{in}^\m{s}}\Delta \hat{c}_\mathrm{in}^\mathrm{s}.
\label{c3ZeroQ}
\end{align}
The resulting concentration $\hat{c}_3=2.2$ in the sieve element is lower than the IC-sugar concentration because a downhill gradient to the SE is essential for diffusion. The water potential $\hat{\Psi}_3$ is calculated with Eq. (\ref{Psi3Dimless}) for zero water outflow $\hat{Q}_\mathrm{out}=0$ as
\begin{align}
\hat{\Psi}_3&=\hat{\Psi}_2+\bar{W}_\m{out}\left[\hat{c}_2-\hat{c}_3\right]
\label{p3ZeroQ}
\end{align}
and we obtain a value of $\hat{\Psi}_3=-0.5$ corresponding to $\Psi_3=-\unit[2.7]{bar}$ and $p_3=\unit[8.3]{bar}$. To generate osmotic water flow into the SE, the xylem pressure has to be larger than $\Psi_3$, i.e., $p_0>\unit[-2.7]{bar}$, which makes it even more difficult for the water to flow directly into the sieve element than into the IC. Thus the water potential in both of the phloem cells (IC and SE) will probably be too high to allow sufficient water intake across the cell membrane from the xylem system.
Furthermore pure diffusion across the IC-SE interface requires that the sugar concentration decreases into the SE [as seen in Eq. (\ref{c3ZeroQ})], which presumably is a disadvantage for efficient sugar translocation. 
In both respects the situation improves, when we allow for water flow through the PD pores in the BSC-IC interface as we show below.

\subsection{Equal concentrations in SE and IC}
\label{secB}
The general case with both diffusion and water flow across both cell interfaces is complicated as seen, for example, from Eq. (\ref{c3Sol}), and one has to deal with many unknown variables, mainly pressures, bulk flows, and the SE concentration. In this subsection we shall therefore treat the special case, where the concentrations in the intermediary cell and sieve element are equal, i.e., $c_2=c_3$, which is likely due to the well connected IC-SE complex. Compared to pure diffusion into the SE this has the advantage, that the concentration of sugar in the phloem sap is higher and therefore the sugar flow will be larger.  As a consequence of the equal concentrations in the phloem, the sugar from the IC will be transported by pure bulk flow from the intermediary cell into the sieve element.  Using (\ref{PhiInHatwithQ}) and  (\ref{PhiOutHatwithQ}), the sugar flows are then expressed as
\begin{align}
\hat{\Phi}_\mathrm{in}&=W_\m{in}^\m{s}\hat{Q}_\m{in}+\frac{H_\m{in}^\m{s}f^\m{s}}{N_\m{in}^\m{s}(\lambda_\m{in}^\m{s})^3}\Delta \hat{c}_\mathrm{in}^\mathrm{s}
\label{PhiIndelc0}\\
\hat{\Phi}_\mathrm{out}&=\bar{W}_\mathrm{out}\hat{Q}_\mathrm{out}\hat{c}_2
\label{PhiOutdelc0}
\end{align} 

\begin{figure}
\centering
\includegraphics{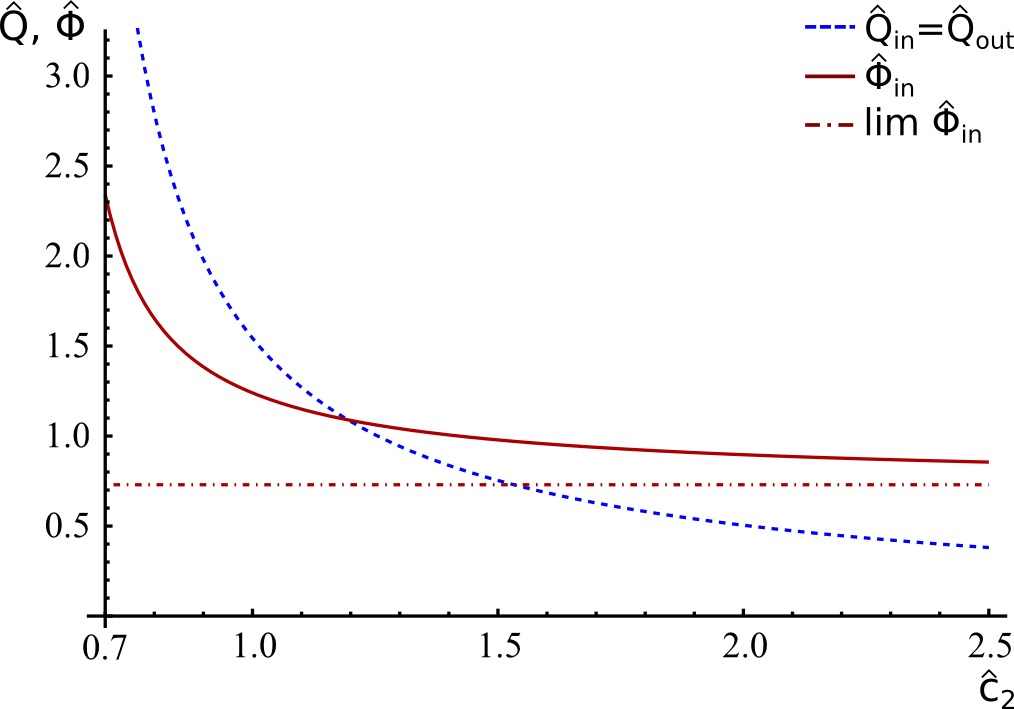}
\caption{\textbf{Water and sugar flow rates $\hat{Q}_\m{in}=\hat{Q}_\m{out}$ (blue, dashed) and $\hat{\Phi}_\m{in}$ (red) as functions of the total sugar concentration $\hat{c}_2$ in the case where the concentrations in IC and SE are equal ($c_2=c_3$).} The flow rates are shown for no trans-membrane flow, i.e., $\hat{Q}_2=0$, only the oligomer concentration $\hat{c}_2^\m{o}$ in the phloem is varied while the sucrose concentration is fixed to $\hat{c}_2^\m{s}=0.7$. The diffusive flow rate into the IC retains its constant value $\lim \hat{\Phi}_\m{in}\propto \Delta \hat{c}_\m{in}^\m{s}$ (dot-dashed), while for an increasing oligomer concentration the advective contribution to the sugar flow decreases with the water flow, which is limited by the conservation laws.}
\label{plotc2Hat_QPhi}
\end{figure}

Using the volume conservation (\ref{eqVolConsDimless}) we can determine the volume flow $\hat{Q}_\m{out}$ and sugar flow $\hat{\Phi}_\m{out}$ from the sugar conservation (\ref{eqMassConsDimless}) with a given trans-membrane flow $\hat{Q}_2$ as functions of the concentration $\hat{c}_2$ in the phloem, i.e.,
\begin{align}
\hat{Q}_\m{out}&=\frac{H_\m{in}^\m{s}f^\m{s}(\lambda_\m{in}^\m{s})^{-3}(N_\m{in}^\m{s})^{-1}\Delta \hat{c}_\m{in}^\m{s}-W_\m{in}^\m{s}\hat{Q}_2}{(x^\m{s}+2x^\m{o})\bar{W}_\m{out}\hat{c}_2-W_\m{in}^\m{s}}
\label{caseBQOutHat}\\
\hat{\Phi}_\m{out}&=\frac{H_\m{in}^\m{s}\bar{W}_\m{out}f^\m{s}(\lambda_\m{in}^\m{s})^{-3}(N_\m{in}^\m{s})^{-1}\Delta \hat{c}_\m{in}^\m{s}\hat{c}_2-W_\m{in}^\m{s}\bar{W}_\m{out}\hat{Q}_2\hat{c}_2}{(x^\m{s}+2x^\m{o})\bar{W}_\m{out}\hat{c}_2-W_\m{in}^\m{s}}
\label{caseBPhiOutHat}
\end{align}
Here the proportions $x^\m{s}$ and $x^\m{o}$ and consequently the average bulk hindrance factor $\bar{W}_\m{out}$ at the IC-SE interface also depend on $\hat{c}_2$. The corresponding inflows are subsequently determined by the conservation laws. The higher we choose the oligomer concentration for a fixed sucrose concentration $\hat{c}_2^\m{s}$ the lower are the resulting flows, approaching the limits 
\begin{align}
\lim_{\hat{c}_2\to \infty}\hat{Q}_\m{out}&=0\\
\lim_{\hat{c}_2\to\infty}{\hat{\Phi}_\m{in}}&=\frac{H_\m{in}^\m{s}f^\m{s}}{(\lambda_\m{in}^\m{s})^{3}N_\m{in}^\m{s}}\Delta \hat{c}_\m{in}^\m{s}-W_\m{in}^\m{s}\hat{Q}_2
\label{limitPhiInc2Large}
\end{align} 
The contribution of the bulk flow to the inflowing sugar current decreases for high IC-concentrations, if there is no runoff of pure water from the IC into the apoplast that would prevent the dilution of the concentrated phloem solution. Since the diffusive contribution stays constant due to the fixed sucrose gradient, the total sugar inflow decreases together with the water flow for a more concentrated phloem solution as seen in Fig. \ref{plotc2Hat_QPhi}.\\

We do not know values for the permeability of the plasma membranes on the loading pathway. Depending on the abundance of {\em aquaporins}, i.e., water-conducting proteins, it can vary by several orders of magnitude between $L_\m{p,2}=\xi_2/A_2=\unit[2\cdot 10^{-14}]{m\,s^{-1}\,Pa^{-1}}$ and $L_\m{p,2}=\unit[10^{-11}]{m\,s^{-1}\,Pa^{-1}}$ as measured by Maurel in plant cells \cite{maurel1997a}. We assume here, however, that the permeability $L_\m{p,2}$ of the IC-plasma membrane is much smaller than the permeabilities $L_\m{p,in(out)}=\xi_\m{in(out)}/A_\m{in(out)}\sim \unit[10^{-12}]{m\,s^{-1}\,Pa^{-1}}$ of the plasmodesmata, and we thus neglect $\hat{Q}_2$ in the following. For this case, Fig. \ref{plotc2Hat_QPhi} shows the behavior of the volume and sugar flows $\hat{Q}_\m{in}=\hat{Q}_\m{out}$ and $\hat{\Phi}_\m{in}$ as functions of $\hat{c}_2$ as in Eqs. (\ref{caseBQOutHat}) and (\ref{caseBPhiOutHat}). For the measured IC concentration of $\hat{c}_2=2.5$ in muskmelon \cite{haritatos1996a} the bulk flow contributes to the sugar inflow only by $15 \%$.
Also for $\hat{Q}_2=0$, we have $\hat{Q}_\m{in}=\hat{Q}_\m{out}$ and the water potentials in the phloem can then be determined as
\begin{align} 
\hat{\Psi}_2&=\hat{p}_2-\hat{c}_2=\hat{\Psi}_1-(\lambda_\m{in}^\m{s})^3\hat{Q}_\mathrm{out}+W_\mathrm{in}^\mathrm{s}\Delta \hat{c}_\m{in}^\m{s}
\label{caseBPsi2Hat}\\
\hat{\Psi}_3&=\hat{\Psi}_2-(\lambda_\m{out}^\m{s})^3 \hat{A}_\m{out}^{-1}\hat{Q}_\mathrm{out}
\label{caseBPsi3Hat}
\end{align}
For the concentrations in \textit{Cucumis melo} and a bundle-sheath pressure of $\unit[1]{bar}$, the resulting values in the phloem are $\hat{\Psi}_2=-0.83$ and $\hat{\Psi}_3=-0.97$ corresponding to dimensional values $\Psi_2=\unit[-4.2]{bar}$ and $\Psi_3=\unit[-4.9]{bar}$ for the potentials and $p_2=\unit[8.3]{bar}$ and $p_3=\unit[7.6]{bar}$ for the hydrostatic pressures. 

\subsection{The loading unit as a part of the phloem}

So far our modeling has not taken into account that the sieve elements are part of the phloem vascular system, and that sap is therefore transported from one sieve element to the next along the phloem vasculature. The pressure drop between the sieve elements needed for this flow is very small compared to the pressure drops across the PDs, which we have been considering so far, since the sieve elements and even the pores in the sieve plates are several orders of magnitude wider. Thus the sieve elements all probably have roughly the same pressures and concentrations. If we also suppose that there is no direct water exchange between the sieve elements and the apoplast,  the sugar and water, which is loaded into the sieve elements, should have those same concentrations. 
The simplified flow in the last subsection, where we assumed equal sugar concentrations in the  IC and SE and thus pure bulk advection through the IC-SE interface, would then be impossible, since it would result in the {\em dilution} of the phloem sap due to the different hindrances of the sugars and the water in the plasmodesmata.
To find an appropriate condition, we denote the sugar flow rate from along the sieve tube (i.e., from one sieve element to the next) by $\Phi_\m{SE}$ and the amount provided by each IC as  $\Delta \Phi_\m{SE}$. If the concentration in the sieve element (of some solute) is $c$, the sugar flow is related to the water flow rate $Q$ simply by $\Phi_\m{SE} = \, Q$ and the condition described above would then amount to $\Delta \Phi_\m{SE} = c \Delta Q = \Phi_\m{out}$, where $ \Phi_\m{out}$ is the flow rate of this particular solute across the IC-SE interface. 
  
With no direct water exchange between the sieve element and the xylem, $\Delta Q=Q_\m{out}$. Thus the conservation laws (\ref{eqMassConsDimless}) and (\ref{eqVolConsDimless}) result in the following equations, where at the IC-SE interface the sucrose and oligomer flux rates are both conserved and can therefore be treated separately, i.e., 
\begin{align}
&\hat{\Phi}_\m{out}^\m{s(o)}=\hat{Q}_\m{out}\hat{c}_3^\m{s(o)} \nonumber \\
\Rightarrow & W_\m{out}^\m{s} \hat{Q}_\m{out}\hat{c}_2^\m{s}+\hat{A}_\m{out}\hat{D}_\m{out}^\m{s}\left[\hat{c}_2^\m{s}-\hat{c}_3^\m{s}\right]=\hat{Q}_\m{out}\hat{c}_3^\m{s}
\label{caseCSucConsSE} \\
\Rightarrow & W_\m{out}^\m{o} \hat{Q}_\m{out}\hat{c}_2^\m{o}+\hat{A}_\m{out}\hat{D}_\m{out}^\m{o}\left[\hat{c}_2^\m{o}-\hat{c}_3^\m{o}\right]=\hat{Q}_\m{out}\hat{c}_3^\m{o}
\label{caseCStaConsSE} \\
&\hat{\Phi}_\m{in}=(x^\m{s}+2x^\m{o})(\hat{\Phi}_\m{out}^\m{s}+\hat{\Phi}_\m{out}^\m{o})\nonumber \\
\Rightarrow & W_\m{in}^\m{s} (\hat{Q}_\m{out}-\hat{Q}_2)+\hat{D}_\m{in}^\m{s}\Delta \hat{c}_\m{in}^\m{s}=(x^\m{s}+2x^\m{o})\hat{Q}_\m{out}(\hat{c}_3^\m{s}+\hat{c}_3^\m{o})
\label{caseCsugConsIC}
\end{align}
Here the dimensionless forms of (\ref{PhiIn}) and (\ref{PhiOut}) of the sugar in and out flow rates are used with $\hat{\Phi}_\m{out}=\hat{\Phi}_\m{out}^\m{s}+\hat{\Phi}_\m{out}^\m{o}$. The average Eq. (\ref{PhiOutDimless}) with $\bar{W}_\m{out}$ and $\bar{D}_\m{out}$ can not be employed here, since the sugar ratios $c_3^\m{s(o)}/c_3$ in the SE are in general not equal to $c_2^\m{s(o)}/c_2=x^\m{s(o)}$ in the IC.
From these equations the SE concentrations $\hat{c}_3^\m{s}$ and $\hat{c}_3^\m{o}$ can be expressed as
\begin{align}
\hat{c}_3^\m{s(o)}=\hat{c}_2^\m{s(o)}\frac{W_\m{out}^\m{s(o)}\hat{Q}_\m{out}+\hat{A}_\m{out}\hat{D}_\m{out}^\m{s(o)}}{\hat{Q}_\m{out}+\hat{A}_\m{out}\hat{D}_\m{out}^\m{s(o)}}.
\label{caseCc3Hat(QOutHat)}
\end{align}
Depending on $\hat{Q}_\m{out}$ the SE concentration $\hat{c}_3=\hat{c}_3^\m{s}+\hat{c}_3^\m{o}$ will take a value between $\hat{c}_2^\m{s} W_\m{out}^\m{s}+\hat{c}_2^\m{o} W_\m{out}^\m{o}$ in the case of a very high advective contribution at the IC-SE interface, and $\hat{c}_2$ for a very high diffusive contribution. The bulk flow $\hat{Q}_\m{out}$ can be determined from (\ref{caseCSucConsSE}), (\ref{caseCStaConsSE}), and (\ref{caseCsugConsIC}) with $\hat{Q}_2=0$. Using the specific values from Table \ref{tabValues}, the resulting SE concentrations in \textit{Cucumis melo} would then be $\hat{c}_3^\m{s}=0.7$ and $\hat{c}_3^\m{o}=1.4$ so that the total SE concentration $\hat{c}_3=2.1$ lies as expected between $\hat{c}_2^\m{s} W_\m{out}^\m{s}+\hat{c}_2^\m{o} W_\m{out}^\m{o}=1.3$ and $\hat{c}_2=2.5$. The bulk contributions to the sugar flow rate at the different interfaces are then calculated with
\begin{align}
\frac{\hat{\Phi}_\m{in}^\m{bulk}}{\hat{\Phi}_\m{in}}&=\frac{W_\m{in}^\m{s}}{(x^\m{s}+2x^\m{o})(\hat{c}_3^\m{s}+\hat{c}_3^\m{o})}=0.14
\label{bulkContIn}\\
\frac{\hat{\Phi}_\m{out}^\m{bulk}}{\hat{\Phi}_\m{out}}&=\frac{W_\m{out}^\m{s}\hat{c}_2^\m{s}+W_\m{out}^\m{o}\hat{c}_2^\m{o}}{\hat{c}_3^\m{s}+\hat{c}_3^\m{o}}=0.62.
\label{bulkContOut}
\end{align}
Thus the advective flow from the intermediary cell into the sieve element in this case contributes about 62 \% to the overall sugar outflow while at the BSC-IC interface the bulk contribution would merely be 14 \%. 
Furthermore the water potentials become $\Psi_2=-\unit[3.9]{bar}$ (IC) and $\Psi_3=-\unit[3.3]{bar}$ (SE) [using Eqs. (\ref{Psi2Dimless}) and (\ref{Psi3Dimless})], and the pressures are $p_2=\unit[8.6]{bar}$ and $p_3=\unit[7.3]{bar}$. So we believe that we have a consistent picture, where all the water necessary for the sap translocation in the phloem is provided together with the sugar through the plasmodesmata with no further need of osmotic water uptake.

\subsection{Diffusion of raffinose}
\label{secraf}
Up to this point, we have treated the oligosaccharides as one species with properties largely determined by stachyose, the one present in largest concentrations. This treatment presumably gives good estimates for the transport rates and water flux, but we still have to account for the fact that raffinose, which is smaller than stachyose, does not diffuse back into the bundle sheath. The transport of raffinose would be given as
\begin{equation}
\hat{\Phi}_\m{in}^\m{r}=\frac12 W^\m{r} \hat{Q}_\m{in}\hat{c}^\mathrm{r} -\frac{H_\m{in}^\m{r}f^\m{r}}{(\lambda_\m{in}^\m{s})^3 N_\m{in}^\m{r}}\hat{c}^\m{r}
\label{difraf}
\end{equation}
where we have used the average raffinose concentration $\hat{c}^\mathrm{r} /2$ between BSC and IC in the advection term. Here we assume that the bulk water flow $\hat{Q}_\m{in}$ is still given by Eq. (\ref{QInHatRewritten}) used above, i.e., 
\begin{align}
\hat{Q}_\m{in}=(\lambda_\m{in}^\m{s})^{-3}\left[\hat{\Psi}_1-\hat{\Psi}_2+W_\m{in}^\m{s}\Delta \hat{c}_\m{in}^\m{s}\right]
\label{QInHatRepeated}
\end{align}
and we investigate whether the bulk flow is sufficient to block the diffusion of raffinose, which would mean that $\hat{\Phi}_\m{in}^\m{r}$ is actually positive.
With the coefficients characterizing the movement of raffinose denoted by the superscript $r$, we get
\begin{align}
\hat{\Phi}_\m{in}^\m{r}\approx\left[\frac{W_\m{in}^\m{r}W_\m{in}^\m{s}}{2}\Delta \hat{c}_\m{in}^\m{s}-\frac{H_\m{in}^\m{r}f^\m{r}}{N_\m{in}^\m{r}}\right]\cdot \frac{\hat{c}^\m{r}}{(\lambda_\m{in}^\m{s})^3}
\label{est}
\end{align}
where we have neglected $\hat{\Psi}_1-\hat{\Psi}_2$, which is typically less than or equal to 0. Using the raffinose radius $r^\m{r}= \unit[0.52]{nm}$ from a 3D-structure model \cite{liesche2013a}, the half-slit width $h_\m{in}=\unit[0.6]{nm}$ as above and the measured free diffusion coefficient $D^\m{r}=\unit[2.15]{m^2\,s^{-1}}$ \cite{dunlop1956a} in cytosol (half of the value in water) with $\Delta \hat{c}_\mathrm{in}^\mathrm{s}=0.3$ we find $\frac12 W_\m{in}^\m{r}W_\m{in}^\m{s}\Delta \hat{c}_\m{in}^\m{s}-\frac{H_\m{in}^\m{r}f^\m{r}}{N_\m{in}^\m{r}}\approx -0.26$ and thus $\hat{\Phi}_\mathrm{in}^\m{r} < 0$ meaning that the bulk flow {\em cannot } block the back diffusion of the intermediate sized raffinose molecules.
\\

Thus, to avoid the diffusion of raffinose back into the bundle sheath we need a half-slit width, which is very close to the radius of the raffinose molecules, denoted by $r^\m{r}$ above. Since these molecules are not spherical, the relevant size depends strongly on how it is defined and/or measured, and thus the hydrodynamic radius of raffinose can vary between values 10\% and 20\% above that of the sucrose molecules. In addition the corresponding value of $\lambda_\m{in}^\m{s} \ge 0.8$ is at the limit (or above) of the range of validity of the hindrance factors, so all in all our results will be somewhat uncertain. Using the value $h_\m{in}=r^\m{r}=\unit[0.52]{nm}$ from 3D-modeling \cite{liesche2013a} gives $\lambda_\m{in}^\m{s} \approx 0.8$ for the sucrose molecules. Using this value in our equations does not change the qualitative features of the solutions obtained above (see Fig. \ref{diffusionPlot}). In this case, using Eq. (\ref{PhiInDiff}), the sugar current would still be larger than the measured value (14 times larger instead of 30 times larger with the half-slit width $h_\mathrm{in}=\unit[0.6]{nm}$). Taking the values $r^\m{s}=\unit[0.52]{nm}$ for the sucrose radius and $r^\m{r}=\unit[0.57]{nm}$ as half-slit width directly from the Einstein relation \cite{liesche2013a} gives us $\lambda_\m{in}^\m{s} \approx 0.9$, and in this case we are above the stated range of validity of $H(\lambda)$. If we use the expressions (\ref{defH}) and (\ref{defW}) we get $H=0.03$ and $W=0.09$. Using again Eq. (\ref{PhiInDiff}) with $f^\m{s}=1$, we obtain
\begin{align}
\hat{\Phi}_\m{in}=\frac{H(\lambda=0.9)}{2 \pi N_\m{A} c_1 (r^\m{s})^3 0.9}\Delta \hat{c}_\mathrm{in}^\mathrm{s}=0.079,
\label{PhiInDiffEinstein}
\end{align}
which is still about three times the measured value 0.025.
To get down to the experimental value we have to decrease the half-slit width below $r^\m{r}$ to $h_\m{in}=\unit[0.54]{nm}$, i.e., $\lambda_\m{in}^\m{s} = 0.96$.

\section{Conclusion}
We have analyzed the feasibility of the polymer trap loading mechanism (active symplasmic loading) in terms of the coupled water  and sugar movement through the plasmodesmata in the cellular interfaces leading from the bundle sheath to the phloem. We used the Kedem-Katchalsky equations and model the pores in the cell interfaces as narrow slits.  This allowed us to compute the membrane coefficients using results on hindered diffusion and convection, and to check whether they can act as efficient filters, allowing sucrose to pass, but not raffinose and stachyose, synthesized in the intermediary cells. Based on the very limited available data for plasmodesmata geometry, sugar concentrations and flux rates, we conclude that this mechanism can in principle function, but, since the difference in size between raffinose and sucrose is only 10-20\%, we are pressing the theories for hindered transport to the limit of (or beyond) their validity. We find that sugar loading is predominantly diffusive across the interface separating the bundle sheath from the phloem. However, the sugar translocation into the sieve tube, where the vascular sugar transport takes place, can be dominated by advection (bulk flow). This allows the plant to build up both the large hydrostatic pressure needed for the vascular sugar transport and the high concentration needed to make this transport efficient. This is possible because the water uptake to the sieve tubes happens directly through the plasmodesmata instead of through aquaporins in the cell membranes of the phloem. Thus, the water in the phloem has to be taken up across the plasma membranes of the pre-phloem pathway, e.g. the bundle sheath cells.
As mentioned earlier, the experimental data available for these plants are very limited. It would be of great importance to have more information on the concentrations and pressures in the cells as well as the diffusivities across the important interfaces. It would also be of importance to extend the analysis of the sugar translocation all the way back to the mesophyll cells, where it is produced. \vspace{0.4cm}\\

{\bf Acknowledgements}\\
\noindent
We are grateful to the Danish Research Council {\em Natur og Univers} for support under the grant 12-126055.

\bibliography{bibliography}

\end{document}